\documentclass[aps,nofootinbib,superscriptaddress, showpacs,preprintnumbers,  nofootinbibt,twocolumn]{revtex4-2}

\usepackage{epsfig}
\usepackage{multirow}
\usepackage{subcaption}
\usepackage{eurosym}
\usepackage{dcolumn}
\usepackage{bm}
\usepackage{enumerate}
\usepackage{float}
\usepackage{epstopdf}
\usepackage{amsmath}
\usepackage{bm}
\usepackage{amsfonts}
\usepackage{amssymb}
\usepackage{graphicx}
\usepackage{alphalph,mathtools}
\usepackage{etoolbox}
\usepackage{color}
\usepackage{booktabs}
\usepackage{hyperref}
\hypersetup{colorlinks,citecolor=blue}
\usepackage{footnote}
\usepackage{makecell,tabularx}

\def\be{\begin{equation}}
\def\ee{\end{equation}}
\def\bea{\begin{eqnarray}}
\def\eea{\end{eqnarray}}

\begin{document}

\title{New Black Hole Solutions in $f(P)$ Gravity and their Thermodynamic Nature}

\author{Aniruddha Ghosh}
 \email{ruddha.g@gmail.com}
 \affiliation{%
 Department of Mathematics, Indian Institute of Engineering Science and Technology, Shibpur, Howrah-711103, India.\\
}%
\author{Ujjal Debnath}%
 \email{ujjaldebnath@gmail.com}
\affiliation{%
 Department of Mathematics, Indian Institute of Engineering Science and Technology, Shibpur, Howrah-711103, India.\\
}%

\begin{abstract}
Black holes are the fascinating objects in the universe.
They represent extreme deformations in spacetime geometry. Here, we construct  $f(P)$ gravity and the first example of static-spherically symmetric black hole solution in $f(P)$ gravity and discuss their thermodynamics. Using the numerical approach and series solution, we discover the solution and demonstrate that it is a generalization of Schwarzschild. The solution is characterized by a single function that satisfies a non-linear fourth-order differential equation. Interestingly, we can analytically calculate the solution's specific heat, Wald entropy, and Hawking temperature as a function of horizon radius. After analyzing the specific heat, we discovered that the black hole is thermodynamically stable over a small horizon radius.\\
\par
\textbf{Keyword:} Black Hole; Modified gravity; Thermodynamic.

\end{abstract}
\maketitle
\section{Introduction}\label{sec1}
Higher-curvature gravity is important in many aspects of high-energy physics. It was motivated by attempts to develop a quantum explanation of the gravitational field. In cosmology, to represent the universe's history uniformly, to account for early time inflation and late time acceleration consistent with observations \cite{bhi1,bhi1.1,bhi1.2,bhi1.3,bhi1.4}. A deeper theoretical motivation is that it occurs naturally in the gravitational effective action of string theory \cite{bhi2} and quantifiable gravitational theory \cite{bhi3}. However, these theories are difficult to examine because of the higher-derivative equations of motion. For instance, in our situation, the equation of motion is a fourth-order non-linear differential equation. The need for a modified theory of gravity arises from several key reasons related to gaps and limitations in our current understanding of gravity\cite{bhinew1,bhinew2,bhinew3}: \\
(i) Unifying Physics: Current theories of gravity, particularly General Relativity (GR), work well at large scales but don't integrate smoothly with quantum mechanics, which governs the very small scales. Physicists are interested in developing a unified theory that encompasses both quantum mechanics and gravity, often referred to as quantum gravity. Modifications to GR might be a step towards this unification.
(ii) Dark Matter and Dark Energy: Observations suggest the existence of dark matter and dark energy(DE)\cite{bhi1.5,bhi1.6,bhi1.7,bhi1.8}, which don't fit neatly into GR. For instance, dark matter(DM) \cite{bhinew4,bhinew5,bhinew5} seems to affect the gravitational dynamics of galaxies, while dark energy is thought to be responsible for the accelerated expansion of the universe. Modifying GR could provide alternative explanations for these phenomena or integrate them into a more comprehensive framework. The literature has several different kinds of dark energy models, including quintessence, tachyon \cite{bhi1.9}, fermionic field \cite{bhi1.10}, and phantom \cite{bhi1.10}. Chromatic dark energy (HDE) \cite{bhi1.12,bhi1.13}, Chaplygin gas \cite{bhi1.11}, novel dark energy (NADE) agegraphic \cite{bhi1.14}
(iii) High-Energy and Extreme Environments: GR is tested extensively under normal conditions, but we have limited experimental data on gravity in extreme environments, such as near black holes or during high-energy cosmic events. Modifications or extensions of GR could potentially explain phenomena observed in these extreme conditions that GR alone cannot.
(iv) Inconsistencies in Observations: Sometimes, observations of cosmic phenomena don’t align perfectly with predictions made by GR. Modifications to the theory might resolve these inconsistencies or better explain certain astrophysical observations. Overall, while General Relativity has been extraordinarily successful in explaining a wide range of gravitational phenomena, modified theories of gravity are explored to address areas where GR might fall short or to provide a more comprehensive understanding of the universe.
\par The process of building modified theories of gravity begins with the addition of a correction term to the Einstein-Hilbert action. Some examples of extra terms in $f(R)$ gravity \cite{bhi4,bhi5,bhi6}, $f(G)$ gravity \cite{bhi7,bhi8}, $f(T)$ \cite{bhi9,bhi10,bhi11,bhi12} or $f(T,T_{G})$ gravity \cite{bhi13,bhi14}, Weyl gravity \cite{bhi15,bhi16}, Lovelock gravity \cite{bhi17,bhi18}, and so on. A significant feature of a higher-order gravity is its linearized spectrum, i.e., On a maximally symmetric background, the metric perturbation propagates just a transverse, massless graviton. Some known examples are  QT gravity \cite{bhi30.3} and certain f(Lovelock) theories\cite{bhi30.1,bhi30.2}.
 One drawback of the curvature-based correction to the Einstein-Hilbert action is that it is dependent on the spacetime dimension, i.e., there are different theories in different dimensions.
 \par A novel category of altered gravity, based on higher-order curvature invariants, was established in \cite{bhi19}. The invariant term was formed by cubic contraction of the three Riemann tensors, and this theory is called Einsteinian cubic gravity (ECG). Some helpful characteristics of this theory(ECG) are as follows: (i) The cubic curvature terms are neither trivial nor topological in four dimensions, (ii) this is the unique cubic theory of gravity that possesses a spectrum identical to the general relativity and (iii) the coupling parameters are dimension
independent.
\par In this paper, we develop vacuum black hole solutions of modified cubic gravity (MCG) \cite{Rev1, rev2} and investigate their thermodynamic nature. The field equations of MCG are complicated fourth-order non-linear differential equations; we have not been able to solve them analytically. However, using a numerical approach and series solution, we can obtain black-hole solutions and compute analytic expression of Hawking temperature T, Wald entropy S, and specific heat C in terms of horizon radius $r_{h}$. Some examples of black holes in $f(R)$ gravity \cite{bhinew7,bhinew8,bhinew9}, $f(G)$ \cite{bhinew10,bhinew11} gravity,$f(T)$ \cite{bhinew12,bhinew13,bhinew14} gravity and so on.
\par The structure of our paper is as follows: In Sec. \ref{sec2}, We construct MCG and its governing vacuum field equations (\ref{10}) in Sec.\ref{sec3}, our main objective is to thoroughly examine the black hole solution for the metric (\ref{11}) that is both static and spherically symmetric and can be characterized by a single function $\psi(r)$. Here, we determine the equation of motion for the two distinct MCG functions, (i) $f(P)=P$ and (ii) $f(P)=P^2$. We also plot numerically the $\psi(r)$ for various values of $\beta$ in Sec. \ref{sec4}, we compute the analytical expressions of Wald entropy, Hawking temperature, and specific heat of the solution, revealing some noteworthy differences from the standard Schwarzschild solution. For instance, some solutions are thermodynamically stable—that is, they have a positive specific heat.  \\\\
\section{FIELD EQUATION OF $f(P)$ GRAVITY}\label{sec2}
Here, we discuss cubic gravity and construct its modified extension, $f(P)$ gravity. The arbitrary contraction of three Riemann tensors constitutes the cubic term in the cubic gravity. The characteristics of cubic gravity are as follows:
(i) It is in the same spectrum as Einstein's gravity, i.e., On a maximally symmetric background, the metric perturbation propagates just a transverse, massless graviton and
(ii) The cubic term is neither trivial nor topological in four dimensions.

Four-dimensional spacetime will be our primary focus. Under these circumstances, a generic nontopological cubic term $P$ would be

 \begin{multline}\label{1}
   P=\beta_1 R_{\mu \phantom{a} \nu \phantom{a}}^{\phantom{a} \rho \phantom{a} \sigma} R_{\rho \phantom{a} \sigma \phantom{a}}^{\phantom{a} \gamma \phantom{a} \delta} R_{\gamma \phantom{a} \delta \phantom{a}}^{\phantom{a} \mu \phantom{a} \nu} +
   \beta_{2} R _{\mu\nu}^{ \rho\sigma} R_{\rho\sigma}^{ \gamma\delta} R_{\gamma\delta}^{\mu\nu}+\\
   \beta_{3}R^{\sigma \gamma}R_{\mu \nu \rho \sigma} R^{\mu \nu \rho \phantom{a}}_{\phantom{a} \phantom{a} \phantom{a} \gamma}+\beta_{4}R R_{\mu \nu \rho \sigma} R^{\mu \nu \rho \sigma}
   +\\
   \beta_{5} R_{\mu\nu\rho\sigma}R^{\mu\rho}R^{\nu\sigma}+\beta_{6} R_{\mu}^{\nu}R_{\nu}^{\rho} R_{\rho}^{\mu}
+\beta_{7}R_{\mu \nu}R^{\mu \nu}R+\beta_{8}R^3 ,
 \end{multline}
 with the conditions
\begin{equation} \label{2}
    \beta_{7}=\frac{1}{12}(3\beta_{1}-24\beta_{3}-16\beta_{3}-48\beta_{4}-5\beta_{5}-9\beta_{6})\
\end{equation}
\begin{equation}\label{3}
  \beta_{8}=\frac{1}{72}(-6\beta_{1}+36\beta_{2}+22\beta_{3}+64\beta_{4}+3\beta_{5}+9\beta_{6})
\end{equation}

Where $\beta_{i}$ dimensionless parameters. Parameter relations (\ref{2}) and (\ref{3}) are provided by the condition (i).

Now if we choose $\beta_{1}=12$, $\beta_{2}=1$, $\beta_{5}=-12$,  $\beta_{6}=8$ and all others  $\beta_{i}$ are $0$ then $P$ would be
\begin{multline}\label{4}
 P=12R_{\mu \phantom{a} \nu \phantom{a}}^{\phantom{a} \rho \phantom{a} \sigma} R_{\rho \phantom{a} \sigma \phantom{a}}^{\phantom{a} \gamma \phantom{a} \delta} R_{\gamma \phantom{a} \delta \phantom{a}}^{\phantom{a} \mu \phantom{a} \nu} +R _{\mu\nu}^{ \rho\sigma} R_{\rho\sigma}^{ \gamma\delta} R_{\gamma\delta}^{\mu\nu}-\\
 12R_{\mu\nu\rho\sigma}R^{\mu\rho}R^{\nu\sigma}+8 R_{\mu}^{\nu}R_{\nu}^{\rho} R_{\rho}^{\mu}
\end{multline}
To create the Einsteinian cubic gravity (ECG), one can utilize the new $P$ in E.q (\ref{4}) as a correction term in the Einstein-Hilbert action.
In the presence of a vacuum, Einsteinian cubic gravity acts as

\begin{equation} \label{5}
      I=\int \sqrt{-g}\left(\frac{1}{2k}[-2 \Lambda+R]+k\lambda P\right)  \ d^4x
    \end{equation}
    With $\lambda$ as the coupling parameter, $k=8\pi G$ is Newton's constant. For additional details, go to \cite{bhfe1,bhfe2}.\

Let's develop $f(P)$ gravity to wrap up this section. Based on the cubic invariant P of (\ref{1}), the theory ensures the previously mentioned requirements by choosing parameters (\ref{2}) and (\ref{3}).
The  action for $f(P)$ characterized by the

\begin{equation} \label{6}
      I=\int \sqrt{-g}\left(\frac{R}{2k}+f(P)\right)  \ d^4x
    \end{equation}
    This theory's field equation is expressed as
    \begin{equation}\label{7}
\varepsilon_{\mu \nu}=P_{\mu \sigma \rho \lambda}R_{\nu}\hspace{0.01cm}^{\sigma \rho  \lambda}-\frac{1}{2}g_{\mu \nu}\mathcal{L}-2\nabla^{\alpha}\nabla^{\beta}P_{\mu \alpha \beta\nu}=0,
    \end{equation}
      Where
   \begin{equation}\label{8}
       P_{i j k l}=\frac{\partial \mathcal{L}} {\partial R^{i j k l}}
          =\frac{1}{2k}g_{i[k}g_{j]l}+f'(P)\frac{\partial P} {\partial R^{i j k l}}
      \end{equation}
\begin{multline}\label{9}
    \frac{\partial P} {\partial R^{i j k l}}=-\beta_{7}  g_{il} g_{jk}R_{ab}R^{ab}+\beta_7 g_{ik} g_{jl} R_{ab} R^{ab}-\frac{1}{2}\beta_{5}R_{il}R_{jk}\\
    +\frac{1}{2}\beta_{5}R_{ik}R_{jl}+\frac{3}{4}\beta_{6}g_{jl}R_{i}^{a}R_{ka}-\frac{3}{4}\beta_{6}g_{il}R_{j}^{a}R_{ka}-\frac{3}{4}\beta_{6}g_{jk}R_{i}^{a}R_{la} +\\
    \frac{3}{4}\beta_{6} g_{ik}R_{j}^{a}R_{la}
+ \frac{1}{2}\beta_{7}g_{jl}R_{ik}R-\\
\frac{1}{2}\beta_{7}g_{jk}R_{il}R-\frac{1}{2}\beta_{7}g_{il}R_{jk}R+\frac{1}{2}\beta_{7}g_{ik}R_{jl}R-\\
3\beta_{8}g_{il}g_{jk}R^2+3\beta_{8}g_{ik}g_{jl}R^2-
\frac{1}{2}\beta_{4}g_{il}g_{jk}R_{ab\gamma \delta}R^{ab \gamma\delta}+\\
\frac{1}{2}\beta_{4}g_{ik}g_{jl}R_{ab\gamma\delta}R^{ab \gamma\delta}+\frac{1}{2}\beta_{5}g_{jl}R^{ab}R_{iakb}+\frac{1}{2}\beta_{3}R_{j}^{a}R_{iakb}-\\
\frac{1}{2}\beta_{5}g_{jk}R^{ab}R_{ialb}+
\frac{1}{2}\beta_{3}R_{l}^{a}R_{ijka}+\\
2\beta_{4}RR_{ijkl}-\frac{1}{2}\beta_{3}R_{k}^{a}R_{ijla}-\frac{1}{2}\beta_{5}g_{li}R^{ab}R_{jakb}-\frac{3}{2}\beta_{1}R_{i}\hspace{0.001cm}^{a}\hspace{0.01cm}_{l}\hspace{0.01cm}^{b}R_{jakb}-\\
\frac{1}{2}\beta_{3}R_{i}\hspace{0.01cm}^{a}R_{jakl}+\frac{1}{2}\beta_{5}g_{ik}R^{ab}R_{jalb}+\frac{3}{2}\beta_{1}R_{i}\hspace{0.01cm}^{a}\hspace{0.01cm}_{k}\hspace{0.01cm}^{b}R_{jalb}+\\
\frac{1}{4}\beta_{3}g_{jl}R_{i}^{ab\gamma}R_{kab\gamma}-\frac{1}{4}\beta_{3}g_{il}R_{j}\hspace{0.01cm}^{ab\gamma}R_{kab\gamma}+3\beta_{2}R_{ij}\hspace{0.01cm}^{ab}R_{klab}-\\
\frac{1}{4}\beta_{3}g_{jk}R_{i}^{ab\gamma}R_{lab\gamma}+
\frac{1}{4}\beta_{3}g_{ik}R_{j}\hspace{0.01cm}^{ab\gamma}R_{lab\gamma}
\end{multline}
We can recover the equation (2.3) described in this paper \cite{bhfe1} by substituting the value $\beta_{1}=12$, $\beta_{2}=1$, $\beta_{5}=-12$,  $\beta_{6}=8$  in equation (\ref{8}), which we verify.
Equation (\ref{7})'s expansion yields the vacuum field equations.
\begin{multline} \label{10}
   G_{\mu \nu}=k\Big( g_{\mu \nu}f(P)- 2f'(P)\frac{\partial P} {\partial R^{\mu \sigma \rho \lambda}}R_{\nu}\hspace{0.01cm}^{\sigma \rho  \lambda}+\\
    4\nabla^{\alpha}\nabla^{\beta}f'(P)\frac{\partial P} {\partial R^{\mu \alpha \beta \nu}}\Big)
\end{multline}
The covariant derivative concerning the spacetime metric $g_{\mu \nu}$ is represented by $\nabla^{\alpha}$.
This reveals the field equations' intricacy.
\section{BLACK HOLE SOLUTIONS IN $f(P)$ GRAVITY}\label{sec3}
This section finds the governing differential equation for the black hole solutions that are static and spherically symmetric and are described by a single function $\psi(r) $, i.e., metrics of the form
\begin{equation} \label{11}
  ds^2=-\psi(r)dt^2+\frac{dr^2}{\psi(r)}+r^2\left(d\theta^2+sin^2\theta d\phi^2 \right)
\end{equation}
Furthermore, as is customary, we assume that the matter's Lagrangian is zero, meaning there is no source.
We can see the intricacy of the computation by looking at equations (\ref{7}),(\ref{8}),(\ref{9}) and (\ref{10}). Because of computational complexity, we perform additional parameter reduction.
\begin{equation} \label{12}
    \beta_{6}=\frac{1}{9}\left(9\beta_{1}-24\beta_{2}-12\beta_{3}-20\beta_{3}-64\beta_{5}+\beta_{5}\right)
\end{equation}
\begin{equation} \label{13}
    \beta_{5}=-\left(6\beta_{1}-24\beta_{2}-12\beta_{3}-32\beta_{4}\right)
\end{equation}
\begin{equation} \label{14}
    \beta_{1}=\frac{1}{3}\left(18\beta_{2}+8\beta_{3}+26\beta_{6}\right
    )
\end{equation}
\begin{equation} \label{15}
    \beta_{4}=-\frac{1}{31}\left(3\beta_{2}+\beta_{3}\right)
\end{equation}
Finally, under the parameter relations (\ref{12})-(\ref{15}) , the cubic invariant $P$ acquires the form
\begin{multline} \label{16}
 P=\frac{1}{186r^6} (3 \beta_{2} + \beta_{3}) \Big(-56 + 56 \psi(r)^3 + 232 r^3 \psi'(r)^3 \\
 -132r^2\psi''(r)+66r^4\psi''(r)^2+7r^6\psi''(r)^3
\\
-186r^2\psi'(r)^2 (-2+r^2\psi''(r) )-12\psi(r)^2 (14-4r\psi'(r)\\
+11r^2\psi''(r))+12r\psi'(r) (4-52r^2\psi''(r)+r^4\psi''(r)^2 )-\\
6\psi(r)(-28+56r^2\psi'(r)^2-44r^2\psi''(r)+11r^4\psi''(r)^2\\
-8r\psi'(r) (-2+13r^2\psi''(r) ) ) \Big)
\end{multline}
Even now, the calculation's complexity remains after reducing the parameters mentioned above.
We defined the parameter
\begin{equation}\label{17}
    \beta=\left(3 \beta_{2} + \beta_{3}\right)
\end{equation}
We can now evaluate the field equations (\ref{10} ) concerning the metric (\ref{11}).
In the following sections, we examine two scenarios to determine the equation of motion: First, $f(P) = P$; second, $f(P) = P^2$. We describe the procedure when $f(P)=P$; the same procedure also applies to $P^2$.
\subsection{\label{sec:levelsub1}SOLUTION FOR $f(P)=P$\protect  \lowercase{}}
The equation of motion for $f(P) = P$ was
\begin{multline} \label{18}
r^4(-1+\psi+r\psi')+\frac{1}{186}k\beta \Big(56+5128\psi^3-304r^3\psi'^3+420r^2\psi''-\\
66r^4\psi''^2+119 r^6\psi''^3+12\psi^2(-850-916r\psi'+155r^2\psi'')+\\
48r^2\psi'^2(35+119r^2\psi''+24r^3\psi''')+12r\psi'(-388+158r^4\psi''^2+\\
r^2\psi''(-332+51r^3\psi'''))+6\psi(836+68r^2\psi'^2+71r^4\psi''^2+12r^6\psi'''^2+\\
8r\psi'(326+56r^2\psi''+9r^3\psi''')+4r^2\psi''(-95+3r^4\psi^4))\Big)=0
\end{multline}
The remaining elements of the modified Einstein equation are either derivatives of this expression, zero, or this expression equivalent. As we can see, the standard Schwarzschild solution is obtained when we enter $\beta=0$. The complexity of the differential equation persists even if we select the most straightforward form, $f(P)=P$. There is no precise analytical way to solve this equation; thus, we address numerical and approximation by series solutions in the following two sub-subsections.
\subsubsection{\label{sec:levelsub1}SERIES SOLUTION }
In this sub-sub-section, we use (\ref{18}) to construct the black-hole solution of (\ref{6}). It is helpful to examine the metric function close to the horizon when studying the black hole solution of this theory. Assuming that the function $\psi$ can be Taylor-expanded around the horizon and is fully regular there. $r=r_{h}$ is horizon surface at which $\psi(r_{h})=0$.It is specifically necessary for the function to be differentiable at $r_{h}$. By Taylor expanding near horizon
\begin{equation} \label{19}
    \psi(r)=\sum_{n=1}^{\infty}a_{n}(r-r_{h})^n
    \end{equation}
Where $a_{n}=\psi^{(n)}(r_{h})/n!$. Also, take note of the fact that surface gravity is $k_{g}=\psi'(r)/2=a_{1}/2$. The idea is to plug this expansion into equation (\ref{18}) and order by order in powers of $(r-r_{h})$ to determine the coefficients $a_{n}$.The associated equations for $n=1$ and $n=2$ are
\begin{multline} \label{20}
    r_{h}^4-r_{h}^5a_{1}+\beta k\Big(-\frac{28}{93}+\frac{776}{93}r_{h}a_{1}-\frac{280}{31}r_{h}^2 a_{1}^2+\\
  \frac{152}{93}r_{h}^3a_{1}^3-\frac{70}{31}r_{h}^22a_{2}+2a_{2}a_{1}r_{h}^3\frac{663}{31}-\frac{952}{31}r_{h}^4a_{1}^22a_{2}+\\
  \frac{11}{31}r_{h}^44 a_{2}^2-\frac{316}{31}r_{h}^5a_{1}4 a_{2}^2-\frac{119}{186}r_{h}^68a_{2}^3-\\
  \frac{192}{31}r_{h}^5a_{1}^26a_{3}-\frac{102}{31}12a_{1} a_{2}a_{3} r_{h}^6\Big)=0,
\end{multline}
\begin{multline} \label{21}
    4r_{h}^3-5r_{h}^4a_{1}-2a_{2}r_{h}^5-a_{1}r_{h}^4+k\beta \Big(-\frac{836}{31}a_{1}+\\
    \frac{776}{3}(2a_{2}r_{h}+a_{1})-\frac{2608}{31}a_{1}^2r_{h}-\frac{280}{31}2a_{1}^2r_{h}-\frac{68}{31}a_{1}^3r_{h}^2+\\
    \frac{152}{93}3r_{h}^2a_{1}^3-\frac{70}{31}(4r_{h}a_{2}+6a_{3}r_{h}^2)+\frac{380}{31}2a_{2}a_{1}r_{h}^2+\\
    \frac{664}{31}(6r_{h}^2a_{1}a_{2}+4a_{2}^2r_{h}^3+6a_{3}a_{1}r_{h}^3)-\frac{448}{31}a_{1}^2r_{h}^32a_{2}\\
    -\frac{952}{31}(8r_{h}^3a_{2}a_{1}^2+6a_{1}^2r_{h}^4a_{3})+\frac{11}{31}16r_{h}^3a_{2}^2-\frac{72}{31}2r_{h}^4a_{1}a_{2}\\
    -\frac{316}{31}(20a_{2}^2r_{h}^4+8a_{2}^3r_{h}^5)-\frac{119}{186}(8a_{2}^36r_{h}^5)-\frac{72}{31}a_{1}^2r_{h}^46a_{3}-\\
    \frac{192}{31}(30a_{1}^2r_{h}^5a_{3}^3+a_{1}^2r_{h}^524a_{4})-\frac{102}{31}(72r_{h}^5a_{1}a_{2}a_{3}+\\
    24a_{2}^2r_{h}^6a_{3}+48a_{4}a_{2}a_{1}r_{h}^6+36a_{3}^2a_{1}r_{h}^6)-\\
    \frac{12}{31}(216a_{3}^3a_{1}r_{h}^6)-\frac{12}{31}24a_{1}a_{4}r_{h}^6\Big)=0
        \end{multline}
        Note that $a_{1}$, $a_{2}$, and $a_{3}$ are included in the first expression. We cannot determine $a_{1}$ using the conventional methods or describe it in these papers \cite{bhfe2,bhs1}. We use a different strategy to gather information about the coefficients, motivated by the paper\cite{bhfe1}. We are expanding each $a_{n}$ regarding $\beta$'s powers.
        \begin{equation} \label{22}
            a_{n}=\sum_{i=0}^{i_{max}}b_{n,i}\beta^i
        \end{equation}
Using this expression in the equation (\ref{20}) and (\ref{21}), we find the terms
$b_{n,i}$.The $b_{n,0}$ terms follow the series expansion of the ordinary Einstein equation around an event horizon,
\begin{equation} \label{23}
    b_{1,0}=\frac{1}{r_{h}}, b_{2,0}=-\frac{1}{r_{h}^2},b_{3,0}=\frac{1}{r_{h}^3},......, b_{n,0}=(-1)^{n+1}\frac{b_{3,0}}{r_{h}^{n-3}}
\end{equation}
\vspace{0.1cm}
\begin{equation}\label{24}
   b_{1,1} = \frac{264k}{31r_{h}^5},
\end{equation}
\begin{equation}\label{25}
\begin{split}
b_{2,1}=&\frac{2\Big(-1440k-648kr_{h}^2+681kr_{h}^5-1580kr_{h}^6}{31r_{h}^{11}} \\
&+\frac{-36kr_{h}^7+72kr_{h}^9\Big)}{31r_{h}^{11}}
\end{split}
\end{equation}
\begin{multline}\label{26}
      b_{1,2}=\frac{24k\Big(-2880k-1296kr_{h}^2+6906kr_{h}^5-3160kr_{h}^6}{961r_{h}^{14}}\\
     \frac{-72 kr_{h}^7+144 kr_{h}^9+93r_{h}^{12}b_{3,1}\Big)}{961r_{h}^{14}}
\end{multline}
As we can observe, there is no nice pattern for the first- and higher-order terms in $\beta$, so we only present a few coefficients here. An intriguing observation is that not all coefficients are necessary. We only need the terms mentioned below in the case of a small $\beta$ limit and a rapidly decreasing factor $\mathcal{O}(\frac{1}{r_{h}})$. By adding up the necessary terms, we can show that
\begin{multline}\label{27}
  a_{1}=\frac{1}{r_{h}} +\frac{264 k \beta}{31r_{h}^5}-
    \frac{75840 k^2 \beta^2}{961 r_{h}^8}- \frac{3456 k^2\beta^2}{961 r_{h}^7} \\
    +\frac{1728 k^2 \beta^2}{961 r_{h}^7} + \frac{3456 k^2  \beta^2}{961 r_{h}^5}
\end{multline}
\begin{multline}\label{28}
 a_{2}=-\frac{1}{r_{h}^2} + \frac{1362 k \beta}{31 r_{h}^6}- \frac{3160 k \beta}{31 r_{h}^5}\\
    -\frac{144 k  \beta}{31 r_{h}^4} + \frac{72 k \beta}{31 r_{h}^4} + \frac{ 144 k\beta}{31 r_{h}^2}
\end{multline}
We have not found all the $b_{n, i}$. Later on, as we will see, all that is required to characterize the black hole thermodynamics is $a_{1}$ and $a_{2}$. The above expressions are true; we can find the relation between the horizon radius  $r_{h}$ and surface gravity $k_{g}$.After figuring out $a_{1}$ and  $a_{2}$, we can calculate the value of $a_{3}$ using (\ref{20}). We can obtain $a_{4}$ as a function of the previous coefficient since it appears in equation (\ref{21}).
Hence, the $n$th equation generally allows us to determine the $a_{n}$. We discover that the series solution can be constructed to allow the black-hole solution to be admitted by theory (\ref{6}).
We can calculate $k_{g}$  as a function of $r_{h}$ using the expression (\ref{27}).
\begin{multline}\label{29}
     k_{g}=\frac{1}{2}\Big(\frac{1}{r_{h}} +\frac{264 k \beta^2}{31r_{h}^5}-
    \frac{75840 k^2 \beta^2}{961 r_{h}^8}-\\
    \frac{3456 k^2\beta^2}{961 r_{h}^7}
    +\frac{1728 k^2 \beta^2}{961 r_{h}^7} + \frac{3456 k^2  \beta^2}{961 r_{h}^5}  \Big)
\end{multline}
\subsubsection{\label{sec:levelsub2} PERTURBATION SOLUTION}
In this subsection, we create the perturbation solution to the differential equation (\ref{18}) for small $\beta$. We consider  the solution of the differential equation (\ref{18}) of the form
\begin{equation}\label{29.0}
  \psi(r) = \psi_0(r) + \beta \psi_1(r) + \beta^2 \psi_2(r)
\end{equation}
 The idea is we put the equation \ref{29.0} in (\ref{18}) and collect the coefficient of $\beta$. Where $\psi(r)$  is the total solution to the problem, $\psi_0(r)$  is the solution to the unperturbed problem and $\psi_1(r),\psi_2(r)$  are corrections due to perturbations of first and second order, respectively. Solving the perturbation problem we get,
\begin{equation}\label{29.1}
    \psi_0(r) = 1 + \frac{C_{1}}{r}
\end{equation}
\begin{equation}\label{29.2}
    \psi_1(r) = -\frac{12 C_{1}^2 \left( \frac{-81}{r^5} - \frac{427 C_{1}}{6 r^6} \right)}{31 r} + \frac{C_{2}}{r}
\end{equation}
\begin{equation}\label{29.3}
\begin{split}
   \psi_2(r) = &\frac{19968768 C_1^3}{961 r^{11}} + \frac{472767840 C_1^4}{10571 r^{12}} + \frac{22809486 C_1^5}{961 r^{13}} \\
   &  +\frac{1944 C_1 C_2}{31 r^6} + \frac{2562 C_1^2 C_2}{31 r^7} + \frac{C_3}{r}
   \end{split}
   \end{equation}
Where $C_{1},C_{2},C_{3}$ are arbitrary constant.We consider that our answer is simplified to the Schwarzschild case when  $\beta \rightarrow 0$.Consequently we get  $C_{1}=-1$.
Therefore, up to the second-order correction term, the perturbed metric is
\begin{equation}\label{29.4}
\begin{split}
 &\psi(r)=1 - \frac{1}{r} + \beta \left( -\frac{12 \left( \frac{427}{6 r^6} - \frac{81}{r^5} \right)}{31 r} + \frac{c_2}{r} \right)\\
  &+ \beta^2 \left( \frac{36 \left( -\frac{3801581}{6 r^{12}} + \frac{13132440}{11 r^{11}} - \frac{554688}{r^{10}} + \frac{13237 c_2}{6 r^6} - \frac{1674 c_2}{r^5} \right)}{961 r} + \frac{c_3}{r} \right)
\end{split}
  \end{equation}
 \subsubsection{\label{sec:levelsub2}NUMERICAL SOLUTION }
The numerical solution is constructed in this subsection and compared to the perturbed solution.
  \begin{figure}[H]
  \centering
    \includegraphics[width=1\linewidth]{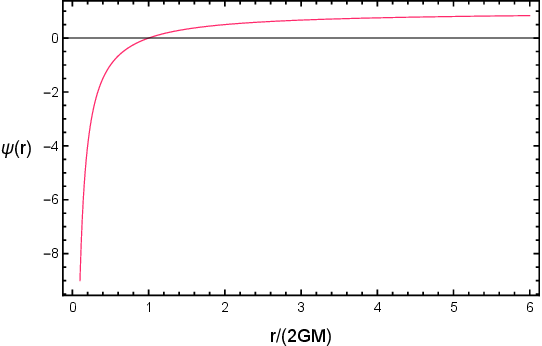}
    \caption{Schwarzschild solution for $\beta=0$}
    \label{f1}
    \end{figure}
\begin{figure}[H]
     \includegraphics[width=90mm,height=60mm]{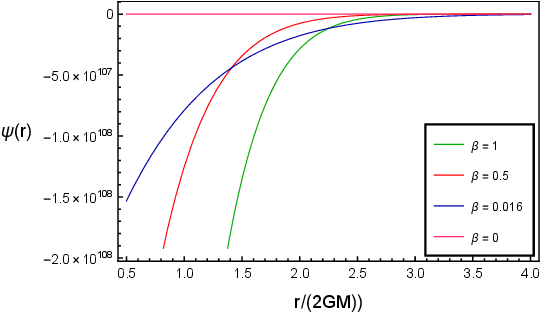}
     \caption{The profile of $\psi(r)$ for various $\beta$ values. The red-pink corresponds to the usual Schwarzchild solution only for $\beta=0$. Schwarzschild solution appears to be a straight line in this instance, but it is not. For several $\beta$ values, the deviation of the other solution is significantly greater than the Schwarzschild, making it appear a straight line.}
     \label{f2}
\end{figure}
\begin{figure}[H]
     \centering
     \includegraphics[width=90mm,height=60mm]{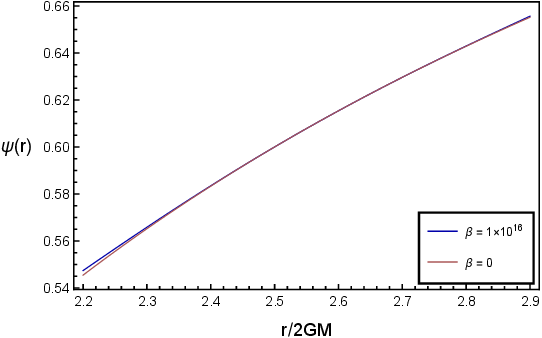}
     \caption{In this figure, we can see that the deviation becomes softer if we select $\beta$ to be very small. }
     \label{f2.0}
     \end{figure}
     \begin{figure}[H]
     \centering
     \includegraphics[width=90mm,height=60mm]{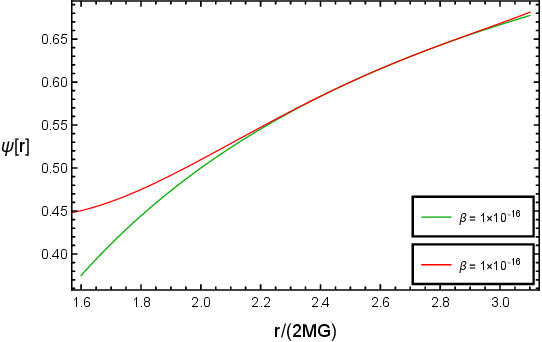}
     \caption{The perturbed and numerical solutions are represented by the green and red curves, respectively. }
     \label{f2.1}
     \end{figure}
     \begin{figure}[H]
     \centering
     \includegraphics[width=90mm,height=60mm]{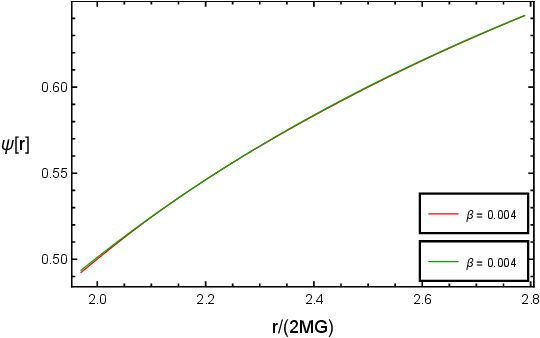}
     \caption{The perturbed and numerical solutions are represented by the green and red curves, respectively.. }
     \label{f2.2}
     \end{figure}
Fig.\ref{f2} shows that the solution converges to a large enough $r$, which approaches the Schwarzschild solution[Fig-\ref{1}]. Furthermore, we can observe that the solution approaches the Schwarzschild solution as $\beta$ approaches zero. We can solve the inner region $\psi<0$ because the horizon is regular. As we can see, the solutions differ noticeably as $r$ approaches the horizon, but they are remarkably similar to Schwarzschild when $r$ is large enough. Once more, note that every solution where $\beta>0$ has a horizon radius larger than the Schwarzschild value, $r_{h}=2MG$. Surprisingly, $\psi(r)$ does not diverge at the origin $r=0$, i.e. there is no metric divergence at the origin. The numerical and perturbed solutions of differential equation (\ref{18}) for various $\beta$ values are plotted in Fig. \ref{f2.1},\ref{f2.2}. Furthermore, we saw that the perturbation solved yields a decent approximation.

 \subsection{\label{sec:levelsub1}SOLUTION FOR $f(P)=P^{2}$\protect  \lowercase{}}
In the preceding sections, we describe the method for Black-hole solution, $f(P)=P$. Similar techniques are used in this section.\\
For the metric of the form (\ref{11}), the equation follows:(see Appendix A). As can be seen, the equation is large and complicated. In this instance, analytical techniques are useless. Even the series solution appears to be an unachievable task. After an extraordinarily lengthy computation,  we obtain relations between the coefficients (In this case, the coefficient of series is $b_{n}$)up to quadratic order. The first equation has 44 terms, while the second has 106 terms. In the case of a small $\beta$ limit and a rapidly decreasing factor $\mathcal{O}(\frac{1}{r_{h}})$ ,the first two terms are;
\begin{equation}\label{32}
    b_{1}=
 \frac{1}{r_{h}}-\frac{111808 k \beta}{2883 r_{h}^{10}} - \frac{5236 k \beta}{2883 r_{h}^5}
\end{equation}
\begin{multline}\label{33}
  b_{2} = -\frac{1}{r_{h}^2}- \frac{4266944 k \beta}{2883 r_{h}^{10}} - \frac{ 823920 k \beta}{961 r_{h}^9}- \\
 \frac{338372 k \beta}{961 r_{h}^8} +\frac{10472 k \beta}{961 r_{h}^6}
\end{multline}
\subsubsection{\label{sec:levelsub1.1} PERTURBATION SOLUTION}
For $f(P)=P^2$ , we obtain  $\psi_0(r)$,$\psi_1(r)$ and $\psi_2(r)$ as follows
\begin{equation}\label{32.1}
    \psi_0(r) = 1 - \frac{1}{r}
\end{equation}
\begin{equation}\label{32.2}
   \psi_0(r) := -\frac{5328 \left( \frac{10627}{15 r^{15}} - \frac{738}{r^{14}} \right)}{961 r} + \frac{C_{1}}{r}
 \end{equation}
 \begin{equation}\label{32.3}
 \begin{split}
 \psi_2(r) :=& -\frac{140409136626135552}{23088025 r^{31}} + \frac{311680223953302528}{26782109 r^{30}} \\
&- \frac{5123168475835392}{923521 r^{29}} + \frac{113241312 C_{1}}{4805 r^{16}} \\
&- \frac{19660320 C_{1}}{961 r^{15}} + \frac{C_{2}}{r}
 \end{split}
\end{equation}
Thus, the metric can be expressed:
\begin{equation}\label{32.4}
\begin{split}
\psi(r)=&1 - \frac{1}{r} - \frac{18873552 \beta}{4805 r^{16}} \\
&+ \frac{3932064 \beta}{961 r^{15}} - \frac{140409136626135552 \beta^2}{23088025 r^{31}}\\
&+ \frac{311680223953302528 \beta^2}{26782109 r^{30}} - \frac{5123168475835392 \beta^2}{923521 r^{29}} \\
&+ \frac{\beta C_{1}}{r} + \frac{113241312 \beta^2 C_{1}}{4805 r^{16}} \\
&- \frac{19660320 \beta^2 C_{1}}{961 r^{15}} + \frac{\beta^2 C_{2}}{r}
\end{split}
\end{equation}

\subsubsection{\label{sec:levelsub2}NUMERICAL SOLUTION }
    \begin{figure}[H]
     \centering
     \includegraphics[width=1\linewidth]{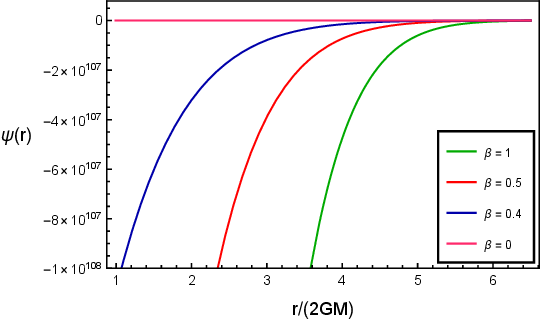}
     \caption{The profile of $\psi(r)$ for various $\beta$ values. The Red-Pink corresponds to the usual Schwarzchild solution only for $\beta=0$. Schwarzschild solution appears to be a straight line, but it is not. For several $\beta$ values, the deviation of the other solution is significantly greater than the Schwarzschild, making it appear a straight line. If we select a value for $\beta$ near zero (Blue line), we can observe that it tends toward Schwarzchild.}
     \label{f3}
\end{figure}
 \begin{figure}[H]
     \centering
     \includegraphics[width=1\linewidth]{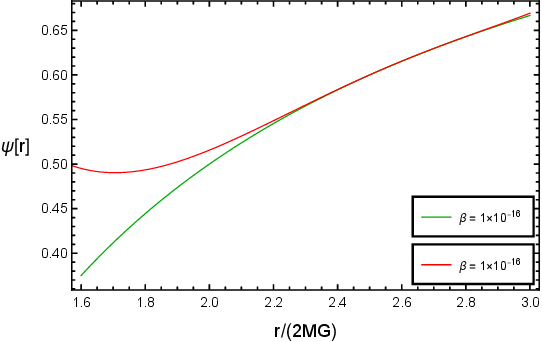}
     \caption{ The perturbed and numerical solutions are represented by the green and red curves, respectively.}
     \label{f3.1}
\end{figure}
As we can see from Fig.\ref{f3}, the curves approach Schwarzchild (Red-Pink) when we choose a value for $\beta$ close to zero. An interesting observation is that the decreasing rate of the solution curve is also observed to increase with an increase in the parameter value. In Fig.(7), we show that the perturbation Solution gives us a good approximation .
 \section{\label{sec4}BLACK HOLE THERMODYNAMICS\protect  \lowercase{}}
 In this section, we go over a few of the thermodynamic characteristics \cite{bhter6,bhter7,bhter8,bhter9} of those solutions. Wald's prescription \cite{bhther1,bhther2} can be utilized to compute the entropy.
 \begin{equation}
     S=-2\pi\oint\ \sqrt{\sigma}P^{abcd} \epsilon_{ab} \epsilon_{cd} \,d^2x
 \end{equation}
 Where $\sigma$ is the determinant of induced metric on the horizon and $\epsilon_{ab}$ is the binormal to the horizon, $\epsilon_{ab}\epsilon^{ab}=-2$. For a metric of the form (\ref{11}) and with a spheric horizon placed at $r=r_{h}$, we determine that the Wald entropy is provided by using the definition of $P^{abcd}$ [in E.q (\ref{9})] given by.
 \begin{multline}\label{35}
    S_{p}=\frac{8 \pi^2 r_{h}^2}{k} - \\
    \frac{ 48 \pi^2 \Big(16 +52 a_{1}^2 r_{h}^2 + 28 a_{2}^2 r_{h}^4 +
    32 a_{1} r_{h} (-3 + 2 a_{2} r_{h}^2)\Big) \beta}{31 r_{h}^2}
 \end{multline}

\begin{widetext}
  \begin{multline}\label{36}
  S_{p^2}= \frac{8 \pi^2 r_{h}^2}{k}-\frac{16 \pi^2 r_{h}^2\Big(16+52 b_{1}^2 r_{h}^2+28b_{2}^2 r_{h}^4+ 32b_{1}r_{h}(-3+2 b_{2}r_{h}^2)\Big)}{961 r_{h}^{10}}\\
  \times  \frac{\Big(-56-264b_{2}r_{h}^2+232b_{1}^3 r_{h}^3+264b_{2}^2 r_{h}^4+56 b_{2}^3 r_{h}^6-168b_{1}^2 r_{h}^2(-2+2b_{2}r_{h}^2)+12b_{1}r_{h}(4-104 b_{2}r_{h}^2+4b_{2}^2 r_{h}^4)\Big)\beta^2}{961 r_{h}^{10}}
  \end{multline}
\end{widetext}
where  $S_{P}$ and $S_{P^2}$  are entropy for $f(P)=P$ and $f(P)=P^2$. Here, we determined the entropy in terms of the horizon radius.

    \begin{figure}[H]
    \centering
    \includegraphics[width=1\linewidth]{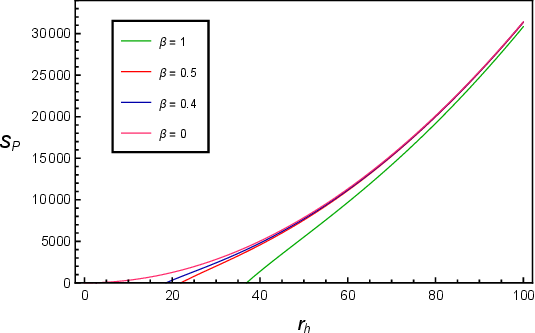}
    \caption{We plot $S_{p}(r_{h})$ for various $\beta$ values.}
    \label{f4}
    \end{figure}
      \begin{figure}[H]
    \centering
    \includegraphics[width=1\linewidth]{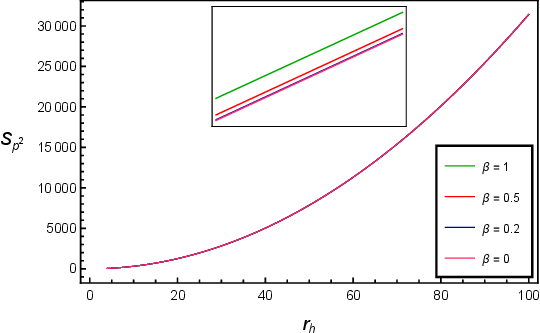}
    \caption{We plot $S_{p^2}(r_{h})$ for various $\beta$ values.}
    \label{f5}
    \end{figure}
    We can observe [from Fig.\ref{f4}, Fig.\ref{f5}]that when the horizon radius increases, the entropy also increases. Additionally, observe that all curves coincide with Schwarzschild when the horizon radius is large enough.
    Here, we observed that the corresponding curve (Blue curve) in Fig.\ref{f5},\ref{f4} coincides with Schwarzschild when we select a value for $\beta$ that is close to zero.

  On the other hand, the Hawking temperature \cite{bhter3} of our solutions can be expressed in terms of the radius as follows:
    \begin{multline}\label{38}
    T_{P}=\frac{\psi'(r_{h})}{4\pi}=\frac{a_{1}}{4\pi}\\
    =\frac{1}{4\pi r_{h}}+\frac{66k\beta}{31\pi r_{h}^5}-\frac{18960 k^2 \beta^2}{961 \pi r_{h}^8}\\
-\frac{432k^2 \beta^2}{961 \pi r_{h}^7}+\frac{864k^2\beta^2}{961 \pi r_{h}^5}
\end{multline}
\begin{equation}\label{39}
 T_{P^2}=\frac{1}{4 \pi r_{h}} -\frac{27952 k \beta}{2883 \pi r_{h}^{10}}-\frac{1309 k \beta}{2883 \pi r_{h}^5}
\end{equation}
Where $T_{P}$ and $T_{P^2}$ are Hawking temperature for $f(P)=P$ and $f(P)=P^2$.
 \begin{figure}[H]
    \centering
    \includegraphics[width=1\linewidth]{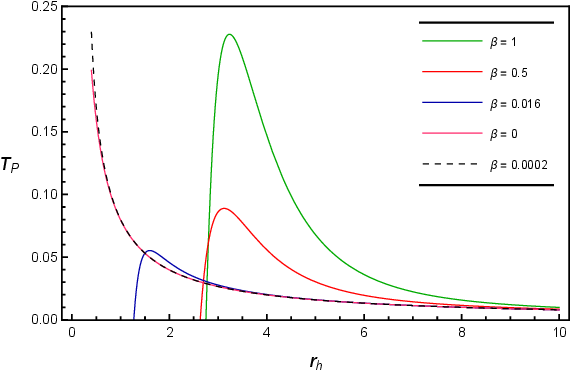}
    \caption{We plot the Hawking temperature as a function of its horizon radius for various  $\beta$ values. When $\beta=0$, the reddish-pink curve corresponds to the standard Schwarzschild solution.}
    \label{f6}
     \end{figure}

    \begin{figure}[H]
    \centering
    \includegraphics[width=1\linewidth]{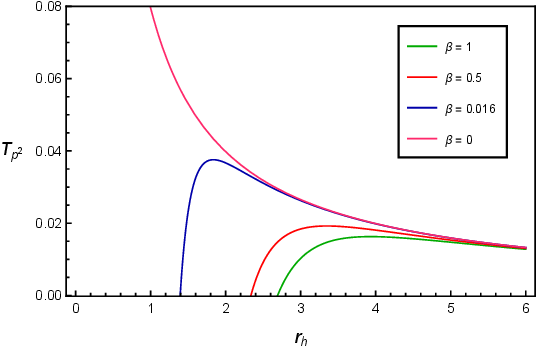}
    \caption{We plot the Hawking temperature as a function of its horizon radius for various  $\beta$ values. When $\beta=0$, the Red-Pink curve corresponds to the standard Schwarzschild solution.}
    \label{f7}
    \end{figure}
 Figure \ref{f6},\ref{f7} illustrates how the temperature increases initially, reaches its maximum value at a certain point, and then begins to decrease. Significantly, the curves for various parameters converge to Schwarzschild for large horizons. We also discovered that this behaviour deviates significantly from Einstein's gravity since Einstein's gravity causes a temperature explosion as $r_{h}$ reaches zero. The temperature behaviour is comparable to that of a standard Reissner-Nordstrom (RN) solution \cite{bhter4}. In that scenario, the temperature reaches its maximum value $T\approx 1/|Q|$  and vanishes as $r_{h} \rightarrow \infty$.
Now, we can calculate specific heat, which we define as
\begin{equation}\label{40}
 C=T\Big(\frac{\partial S}{\partial T}\Big)_{M}
\end{equation}
Everything required to calculate equation (\ref{40})  is already in our possession.
Now that we have entered all values into Equation (\ref{40}), we have the specific heat for $C_P, C_{p^2}$ corresponding to $f(P)=P$ and $f(P)=P^2$.

    \begin{figure}[H]
    \centering
    \includegraphics[width=1\linewidth]{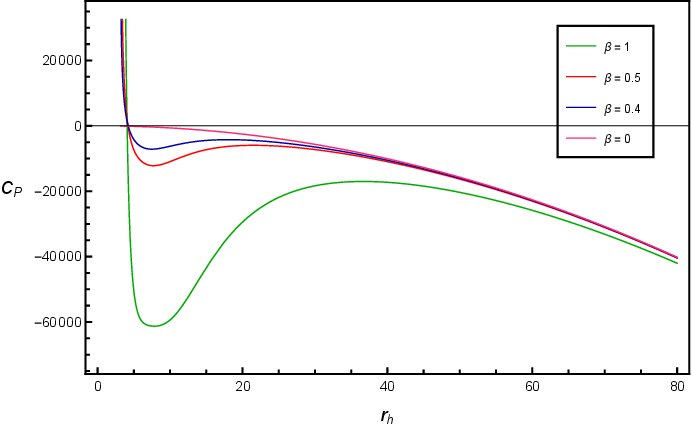}
    \caption{We plot the specific heat as a function of its horizon radius for various  $\beta$ values. When $\beta=0$, the Red-pink curve corresponds to the standard Schwarzschild solution.}
    \label{f8}
    \end{figure}
    \begin{figure}[H]
    \centering
    \includegraphics[width=1\linewidth]{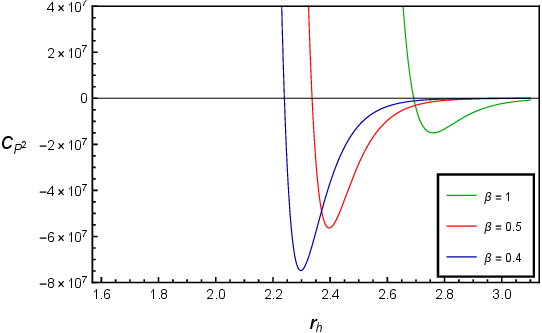}
    \caption{Smaller horizons correlate with positive specific heat.
}
    \label{f9}
    \end{figure}
    The solution is thermodynamically stable if its specific heat is positive. Figure \ref{f8},\ref{f9} shows that positive specific heat is produced as the horizon radius decreases. i.e. it is thermodynamically stable: it contains positive specific heat.
    This type of solution is distinct from the standard Schwarzschild solution, which has $C(r_{h})<0$ for all values of $r_{h}$. This situation will be addressed in \cite{bhfe2,bhter5}, where a black hole is shown to become stable for a small horizon.

\section{\label{sec:level4}CONCLUSIONS\protect  \lowercase{}}
In this study, we constructed the  $f(P)$ gravity and examined the thermodynamic nature of the black hole solution. Also, we demonstrate that MCG reduced it to cubic gravity (ECG) for $\beta_{i}$ is selected appropriately, equations (\ref{1}),(\ref{10}) and (\ref{9}) are reduced to cubic gravity (ECG) and perfectly true. The invariant $P$, which is created using a certain cubic contraction three Riemann tensor, is the foundation of cubic gravity, such that the theory (i) is in the same spectrum as Einstein's gravity, i.e., On a maximally symmetric background, the metric perturbation propagates just a transverse, massless graviton and (ii) The cubic term is neither trivial nor topological in four dimensions. These characteristics make MCG a suitable candidate for modified gravity. Also, those equations are established for the first time.
\vspace{3mm}
\par (i) We have demonstrated that the theory admits a vacuum black hole solution that is both static and spherically symmetric, with a single function $\psi(r)$ that is defined by a non-linear differential equation (\ref{18}),(\ref{sec:sample: appendix}). Remarkably, we can demonstrate that the solution is a generalization of the Schwarzschild black hole using \textcolor{blue}{perturbation} and numerical solutions. Furthermore, we create an approximation of the solutions given by equations (\ref{29.4}) and (\ref{32.4}). In Fig.\ref{f2.1},\ref{f2.2} and \ref{f3.1}, we analyzed the numerical and perturbation solutions of the Black-Hole metric.
\vspace{3mm}
\par (ii)We determine the coefficients $a_{1}$, $a_{2}$, $b_{1}$, and $b_{2}$ for $f(p)=p,p^{2}$ respectively using the series solution approach.To investigate the thermodynamical characteristics of black holes, we require this coefficient.
\vspace{3mm}
\par (iii)  By utilizing those findings, we have examined black holes' thermodynamic properties. we study how Wald's entropy and Hawking temperature change with horizon radius [Fig.\ref{f4},\ref{f5},\ref{f6},\ref{f7}]. One noteworthy aspect of the solutions developed here is the presence of a stable black hole. This type of solution is distinct from the standard Schwarzschild solution in that it is thermodynamically stable: it contains positive specific heat.There are two potential solutions that we have identified  (Fig.\ref{f8},\ref{f9}): (i) a large black hole with $C(r_{h})<0$ and (ii) a smaller black hole with $C(r_{h}) > 0$.This kind of solution has stability with a reduced horizon as one of its properties. Some examples are addressed in \cite{bhfe2,bhter5}, where a black hole is shown to become stable for a small horizon.
\vspace{3mm}
\par (iv) We created the black hole solution for $f(p)=p,p^{2}$ in this study. The differential equations described by the metric function are complex non-linear fourth-order ODEs. There is no known analytical solution for these kinds of problems. Thus, we decided to use (i) a Series solution and (ii) a perturbation solution to solve the differential equation approximately. In the perturbation solution, we obtain the approximate metric function up to the second-order correction term of $\beta$.
\vspace{3mm}
\par (iv) While these black holes share features and thermodynamic behaviour with their Einstein and Lovelock gravity counterparts, they also exhibit some unique qualities. One possibility is the presence of black holes with a small radius, depending on the value of parameter $\beta$. There is still a lot of work to be done for MCG in the future. In the context of black hole thermodynamics, it would be intriguing to observe the impact of including the cubic Lovelock and Gauss-Bonnet components. Adding matter sources, such as a Maxwell field, might also give the thermodynamic behaviour additional structure.
  \\\\
\section*{Acknowledgements}

AG is thankful to IIEST, Shibpur, India, for providing an Institute Fellowship (JRF).
\appendix
 \section{EQUATION OF MOTION FOR $f(P)=P^2$ }
 \label{sec:sample: appendix}
 Substituting the metric (\ref{11}) in equation (\ref{10}) yields the following equation.
\begin{widetext}
  \begin{multline}
r^{10}(r \psi '+\psi -1) +\frac{k \beta}{34596}\Big(-3136+5222336\psi^6-5288512 r^6 \psi'^6-47040 r^2 \psi''-86064 r^4 \psi''^2+42112r^6 \psi''^3-31740 r^8 \psi''^4+15708 r^{10} \psi''^5\\
+1715 r^{12}\psi''^6-384 r^5\psi'^5(7708-22508r^2\psi''+813 r^3\psi''')-192r^4\psi'^4(-3730-439 r^4 \psi''^2+9324 r^3 \psi'''+r^2 \psi''(-33926+1923 r^3 \psi'''))+\\
12r\psi'(3076r^{10}\psi''^5+64(1939-495 r^3 \psi''')+16r^2\psi''(17510+1623r^3\psi''')+8 r^6\psi''^3(-7141+1683r^3 \psi''')+r^8\psi''^4(-35452+1785r^3\psi''')\\
-4r^4\psi''^2(43880+3789r^3\psi'''))+24r^2\psi'^2(-5935r^8\psi''^4+8r^6\psi''^3(-8747+489r^3\psi''')-64(1303+957r^3\psi''')\\
+\psi''(472880r^2-39024r^5\psi''')+
\psi''^2(281388r^4-31752r^7\psi'''))-16r^3\psi'^3(319136+20606r^6\psi''^3-69876r^3\psi'''\\
+12r^2\psi''(26462+9801r^3\psi''')
+\psi''^2(68064r^4+7479r^7\psi'''))-192\psi^5(135982+51484r\psi'+37003r^2\psi''-10584r^3\psi'''+792r^4\psi'''')\\
-48\psi^4(387176r^2\psi'^2+259793r^4\psi''^2+4(-271915+42336r^3\psi'''+750r^6\psi'''^2-3168r^4\psi'''')\\
+4r^2\psi''(-147767-19440r^3\psi'''+750r^4\psi'''')-8r\psi'(106846+144961r^2\psi''-34254r^3\psi'''+2664r^4\psi''''))\\
-12\psi^2(2976032r^4\psi'^4+37685r^8\psi''^4+16(-135835+42336r^3\psi'''+2250r^6\psi'''^2-3168r^4\psi'''')\\
-32r^2\psi''(73271+29160r^3\psi'''+45r^6\psi'''^2-1125r^4\psi'''')+24r^4\psi''^2(130793+3456r^3\psi'''+198r^6\psi'''^2-30r^4\psi'''')\\
+176r^6\psi''^3(-2642-162r^3\psi'''+9r^4\psi'''')-16r^3\psi'^3(-566108+354424r^2\psi''-67581r^3\psi'''+4824r^4\psi'''')\\
+48r^2\psi'^2(107218+50569r^4\psi''^2+82362r^3\psi'''+378r^6\psi'''^2-6060r^4\psi''''+r^2\psi''(-398272-13527r^3\psi'''+378r^4\psi''''))\\
-4\psi'(103426r^7\psi''^3-32r(-31559+26433r^3\psi'''+441r^6\psi'''^2-1998r^4\psi'''')+12r^3\psi''(307432+36663r^3\psi'''+300r^6\psi'''^2-1176r^4\psi'''')\\
+3r^5\psi''^2(-549944-12375r^3\psi'''+600r^4\psi'''')))-32\psi^3(-1618540r^3\psi'^3+86528r^6\psi''^3\\
-4(-407750+95256r^3\psi'''+3375r^6\psi'''^2-7128r^4\psi'''')+6r^2\psi''(221038+58320r^3\psi'''+45r^6\psi'''^2-2250r^4\psi'''')
\\
+3r^4\psi''^2(-390586-5184r^3\psi'''+45r^4\psi'''')+6r^2\psi'^2(-300806+567853r^2\psi''-119715r^3\psi'''+9090r^4\psi'''')-6\psi'(200744r^5\psi''^2\\
+4r(-84982+51876r^3\psi'''+441r^6\psi'''^2-3996r^4\psi'''')+r^3\psi''(-84982-54183r^3\psi'''+1764r^4\psi'''')))\\
+12\psi(103788r^5\psi'^5-3445r^10\psi''^5-16r^2\psi''(36023+19440r^3\psi'''+45r^6\psi'''^2-750r^4\psi'''')\\
+32(-13559+5292r^3\psi'''+375r^6\psi'''^2-396r^4\psi'''')+8r^4\psi''^2(132586+5184r^3\psi'''+594r^6\psi'''^2-45r^4\psi'''')\\
+2r^8\psi''^4(20165+324r^3\psi'''+105r^4\psi'''')+8r^6\psi''^3(-297203564r^3\psi'''+105r^6\psi'''^2+198r^4\psi'''')\\
-32r^4\psi'^4(-91136+67309r^2\psi''-12645r^3\psi'''+378r^4\psi''')+4\psi'(16657r^9\psi''^4-4r^3\psi''(342452+57429r^3\psi'''+900r^6\psi'''^2-1764r^4\psi'''')\\
+16r(-20627+18612r^3\psi'''+441r^6\psi'''^2-1332r^4\psi'''')+6r^5\psi''^2(148456+6819r^3\psi'''+36r^6\psi'''^2-300r^4\psi'''')\\
+2r^7\psi''^3(-44572-3285r^3\psi'''+36r^4\psi''''))
+8r^3\psi'^3(645892+126619r^4\psi''^2+123516r^3\psi'''+564r^6\psi'''^2-9648r^4\psi''''\\
+\psi''(-655924r^2-2322r^5\psi'''+564r^6\psi''''))-4r^2\psi'^2(14803r^6\psi''^3-4(128066+135027r^3\psi'''+1134r^6\psi'''^2-9090r^4\psi'''')\\
+12r^2\psi''(228691+11901r^3\psi'''+63r^6\psi'''^2-378r^4\psi'''')+\psi''^2(-466134r^4-30753r^7\psi'''+378r^8\psi'''')))\Big)=0
\end{multline}
\end{widetext}

\bibliographystyle{naturemag}
\bibliography{bibliography}
\end{document}